\documentclass{article}

\usepackage{float}
\usepackage{graphicx}

\usepackage[final]{neurips_2022}

\usepackage[utf8]{inputenc} 
\usepackage[T1]{fontenc}    
\usepackage{hyperref}       
\usepackage{url}            
\usepackage{booktabs}       
\usepackage{amsfonts}       
\usepackage{nicefrac}       
\usepackage{microtype}      
\usepackage{xcolor}         

\usepackage{color}
\usepackage{biblatex}

\title{Exploring evolution-aware \& -free protein language models as protein function predictors}

\author{%
  Mingyang Hu\footnotemark[1] \\
  Westlake University\\
  \texttt{humingyang@westlake.edu.cn} \\
   \And
   Fajie Yuan\footnotemark[1] \footnotemark[2] \\
   Westlake University\\
  \texttt{yuanfajie@westlake.edu.cn} \\
   \AND
   Kevin K. Yang \\
   Microsoft Research New England \\
   \texttt{yang.kevin@microsoft.com} \\
   \And
   Fusong Ju \\
   Microsoft Research Asia \\
   \texttt{fusongju@microsoft.com} \\
   \And
   Jin Su \\
  Westlake University\\
  \texttt{sujin@westlake.edu.cn} \\
   \AND
   Hui Wang \\
  Westlake University\\
  \texttt{wanghui@westlake.edu.cn} \\
   \And
   Fei Yang \\
   Zhejiang Lab \\
   yangf@zhejianglab.com \\
   \And
   Qiuyang Ding \\
  Westlake University\\
  \texttt{dingqiuyang@westlake.edu.cn} \\
}

\begin{document}

\maketitle
\renewcommand{\thefootnote}{\fnsymbol{footnote}}
\footnotetext[1]{Equal Contribution}
\footnotetext[2]{Corresponding author.  Fajie designed the idea and led the research. Mingyang performed the research and led all experiments. Kevin provided important guidance for this research and worked on a part of paper writing.
Jin performed experiments for the MSA generation. }
\begin{abstract}

Large-scale Protein Language Models (PLMs) have improved performance in protein prediction tasks, ranging from 3D structure prediction to various function predictions. In particular, AlphaFold~\cite{jumper2021highly}, a ground-breaking AI system, could potentially reshape structural biology. However, the utility of the PLM module in AlphaFold, Evoformer, has not been explored beyond structure prediction. In this paper, we  investigate the representation ability of three 
popular PLMs: ESM-1b (single sequence)~\cite{rives2021biological}, MSA-Transformer (multiple sequence alignment)~\cite{rao2021msa} and Evoformer (structural), with a special focus on Evoformer. Specifically, we aim to answer the following key questions: (\romannumeral1) Does the Evoformer trained as part of AlphaFold produce representations amenable to predicting protein function?  (\romannumeral2) If yes, can Evoformer replace ESM-1b and MSA-Transformer? (\romannumeral3) How much do these 
 PLMs rely on evolution-related protein data? In this regard, are they complementary to each other?
  We compare these models by empirical study along with new insights and conclusions.
All code and datasets for reproducibility are available at \textcolor{blue}{\url{https://github.com/elttaes/Revisiting-PLMs}}.

\end{abstract}

\section{Introduction}
Proteins perform the majority of biological activities. 
It is, therefore, crucial to decipher the mechanisms underlying their structural and functional properties. The canonical \emph{sequence-structure-function} relationship enables the success of sequence-based machine learning methods that infer protein structure and function from amino acid (AA) sequence. 
Large-scale protein language models (PLMs) with self-supervised pretraining on tens of millions to billions of proteins  (PLMs)~\cite{rao2019evaluating,rives2021biological,elnaggar2020prottrans,bepler2021learning} are the curent state-of-the-art in predicting function and fitness from sequences. 

Meanwhile, AlphaFold~\cite{jumper2021highly}, trained on experimental 3D protein structures from the Protein Data Bank (PDB)~\cite{sussman1998protein} can approach the resolution of experimental structures for most protein sequences. Its multiple sequence alignment representation module, Evoformer, combines new deep learning machinery, a PLM residue reconstruction task, and structural supervision in the form of a distogram. Like MSA-Transformer~\cite{rao2021msa}, Evoformer takes a family of evolutionarily-related and aligned protein sequences as input, in contrast to PLMs such as ESM-1b~\cite{rives2021biological} and TAPE~\cite{rao2019evaluating}, which only take individual protein sequences. Thus, for short, we refer to the former two models as evolution-aware PLMs and the latter two as evolution-free PLMs.

Despite the remarkable success of AlphaFold in predicting structure from sequence, it is unknown whether its Evoformer module can be applied to other problems, in particular predicting protein function and fitness. Deciphering AlphaFold rather than treating it as a black box is beneficial to both AI and biology communities. 
Therefore, we attempt to answer the following key questions.

\textbf{Q(\romannumeral1): Does the Evoformer trained during AlphaFold training learn general-purpose protein representations that can be used for various function prediction tasks?} Unlike ESM-1b and MSA-Transformer, Evoformer is trained with supervision from 3D structures. In addition, the second part of AlphaFold, the Structure Module, built on top of the 48 Evoformer blocks, is much more complex and deeper than the traditional (linear) classification head used in ESM-1b and MSA-Transformer. These differences make the function representation ability of Evoformer an open question.


\textbf{Q(\romannumeral2): If Evoformer's representation is general, does it outperform ESM-1b and MSA-Transformer on downstream tasks?} While these three models are trained with different parameter sizes and datasets (see Table~\ref{paramsCompare}),  we believe the comparison results are still valuable because they are currently the most advanced PLMs. Training these large models from scratch is out of the reach of most academic research groups due to the compute and cost involved.  
In addition, we  also investigate the utility of MSA-Transformer on function prediction tasks.


\textbf{Q(\romannumeral3): How much does the performance of evolution-aware PLMs rely on the input MSAs? Can evolution-free PLMs assist evolution-aware PLMs in terms of MSA construction?}
Both Evoformer and MSA-Transformer take sets of aligned sequences (MSAs) as input. We investigate the effect of MSA quality and depth on function prediction. On the other side, ESM-1b can simply formulate MSA construction as the remote homology detection task. Then, it is interesting to know whether ESM1b-constructed MSAs can be used as inputs to Evoformer and MSA-Transformer.



We address the above questions through comprehensive empirical studies on a variety of structure and function prediction tasks. Note although AlphaFold (i.e. Evoformer + Structure Module) can accurately predict protein 3D structures, we investigate the ability of Evoformer to perform other structure prediction tasks, such as secondary structure prediction and contact map prediction. 
We make the following key observations:

(\romannumeral1)
The AlphaFold-trained Evoformer produces representations that are useful for both structure and function prediction, as shown on two structure prediction tasks, two function annotation tasks~\cite{zhou2019cafa}, and two fitness score prediction tasks~\cite{rao2019evaluating}.

(\romannumeral2)
Evoformer representations are useful for both protein-level and residue-level prediction tasks.

(\romannumeral3)
Evoformer is superior to ESM-1b and MSA-Transformer for structure prediction and novel miniprotein stability prediction, but inferior to ESM-1b on other function prediction tasks. It performs poorly for zero-shot fitness prediction tasks~\cite{meier2021language} compared to ESM-1b and MSA-Transformer.

(\romannumeral4)
Evolution-aware PLMs are superior to evolution-free ESM-1b model only in the structure prediction tasks, but in general, are worse than ESM-1b in most function prediction tasks.

(\romannumeral5)
Like structure prediction, evolution-aware PLMs are also sensitive to the amount of MSAs when predicting protein functions. In addition, their performance using ESM-1b-constructed MSAs as input is comparable to the performance using MSAs generated by Jackhmmer~\cite{johnson2010hidden} or HHblits~\cite{HHblits}.

\section{Related Work}
\paragraph{Protein language models}

\begin{figure}[t]
  \centering
  \includegraphics[width = 0.9\textwidth]{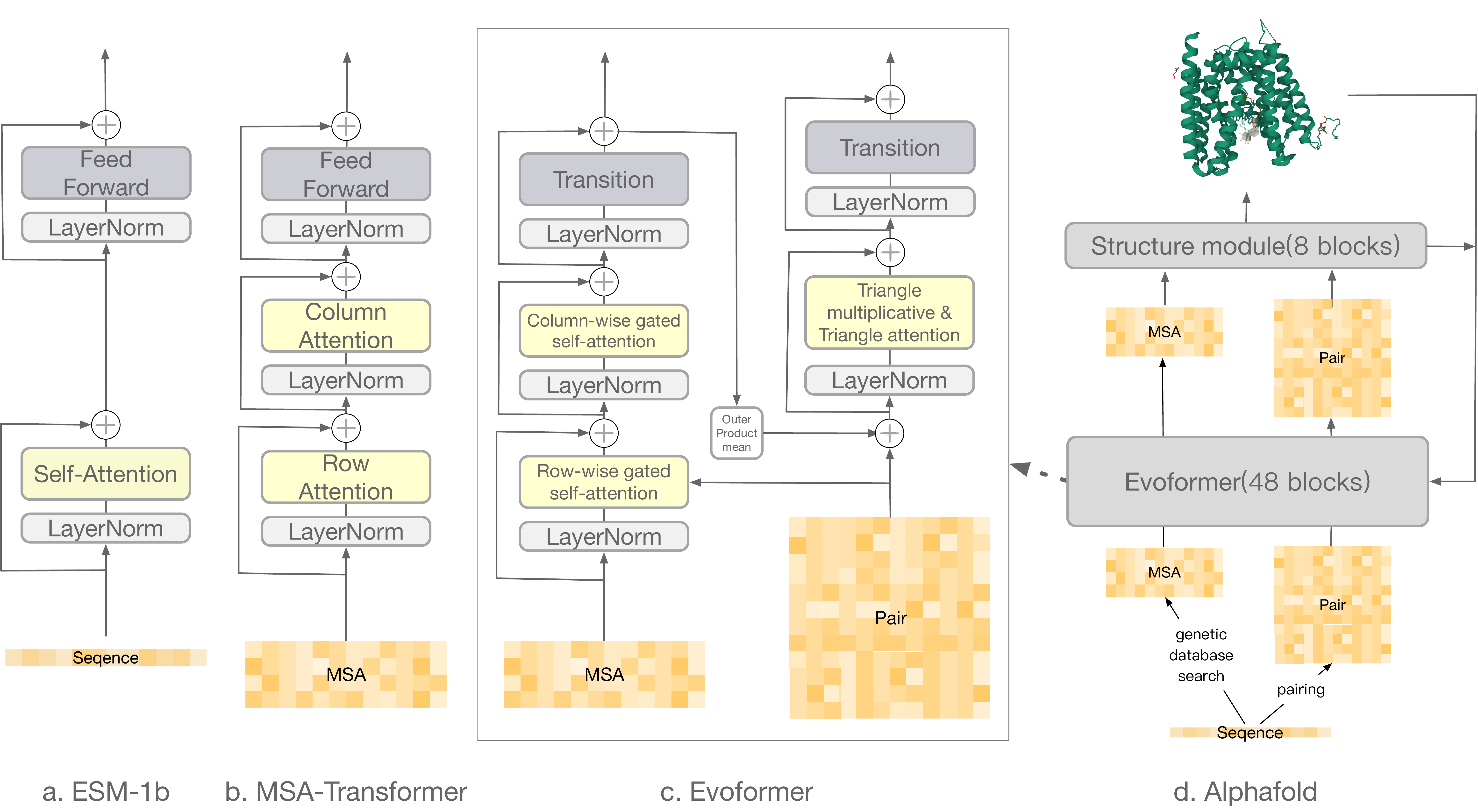}
  \caption{Core modules of the three PLMs.}
  \label{network}
\end{figure}

 The volume of protein data has exploded over the last decade with the advancement of new DNA sequencing technologies. 
Early work learned protein representations with LSTMs~\cite{heinzinger2019modeling,bepler2019learning,alley2019unified}. 
Recently, with the advent of large transformer models in natural language processing (NLP), large PLMs using the Transformer~\cite{vaswani2017attention} architecture and BERT~\cite{devlin2018bert} denoising task have been widely adopted~\cite{rives2021biological,rao2019evaluating,elnaggar2020prottrans}.\footnote{Note that the bioRxiv version of ~\cite{rives2021biological} is  concurrent with ~\cite{heinzinger2019modeling} and ~\cite{alley2019unified}. }
PLMs trained on large sequence databases have been successfully applied for various protein related tasks, including secondary structure prediction~\cite{rao2019evaluating,heinzinger2019modeling,rives2021biological,elnaggar2020prottrans},  contact prediction~\cite{rao2020transformer,rives2021biological}, 3D structure prediction~\cite{jumper2021highly,baek2021accurate},  annotation prediction~\cite{bileschi2019annotate}, signal peptide prediction~\cite{teufel2021signalp}, intracellular localization prediction~\cite{thumuluri2022deeploc}, protein-protein interaction prediction~\cite{evans2021protein}, and fitness prediction~\cite{alley2019unified,dallago2021flip,hie2021learning,hie2022evolocity,meier2021language}.
 ESM-1b~\cite{rives2021biological} found that residue-residue contacts can be recovered from the learned representation, identifying the close relationship between Transformer attention and biological features.
Following this, ~\cite{vig2020bertology} and~\cite{rao2020transformer} further studied the interpretability of the attention map as contact map. 
ProtTrans~\cite{elnaggar2020prottrans} benchmarked a variety of BERT-like PLMs, including TAPE, ESM-1b, ProtTrans, and MSA-Transformer. 
Since ProtTrans with the BERT architecture does not exceed  ESM-1b, we choose to compare  ESM-1b,  MSA-Transformer, and Evoformer for this study. Table~\ref{paramsCompare} and Figure~\ref{network} provide more details about the three models.

 \begin{table}[t]
  \caption{Model descriptions. `Para.', `M', \& `seqs' denotes parameters, million, \& protein sequences.}
  \label{paramsCompare}
  \centering
  \begin{tabular}{lrrrr}
    \toprule
    \cmidrule(r){1-2}
    Model     & Embedding    & Layers & Para. &Training database \\
    \midrule
    ESM-1b & 1280  & 33 &650M&UniRef50 (27M seqs)   \\
    MSA-Transformer     & 768 & 12 &100M&UniRef50 (26M MSAs)    \\
    Evoformer (No Template)     & 256 \& 128 & 48 &88M& PDB (190K structures  +  MSAs) \\
    \bottomrule
  \end{tabular}
\end{table}

\paragraph{Structure, function, and fitness prediction}
A protein's primary structure (i.e. AA sequence) determines its 3D structure, which in turn determines its functional properties. This relationship underlies the success of PLMs, which infer protein structure and function from the raw sequence. Secondary structure (SS) prediction~\cite{sonderby2014protein} is an easier task than predicting the 2D contact map~\cite{wang2017accurate} and 3D structure~\cite{jumper2021highly}.
MSA-Transformer and ESM-1b can make accurate unsupervised contact predictions, which are an important input for the 3D structure generation.

\begin{figure}
  \centering
  \includegraphics[width = 1.0\textwidth]{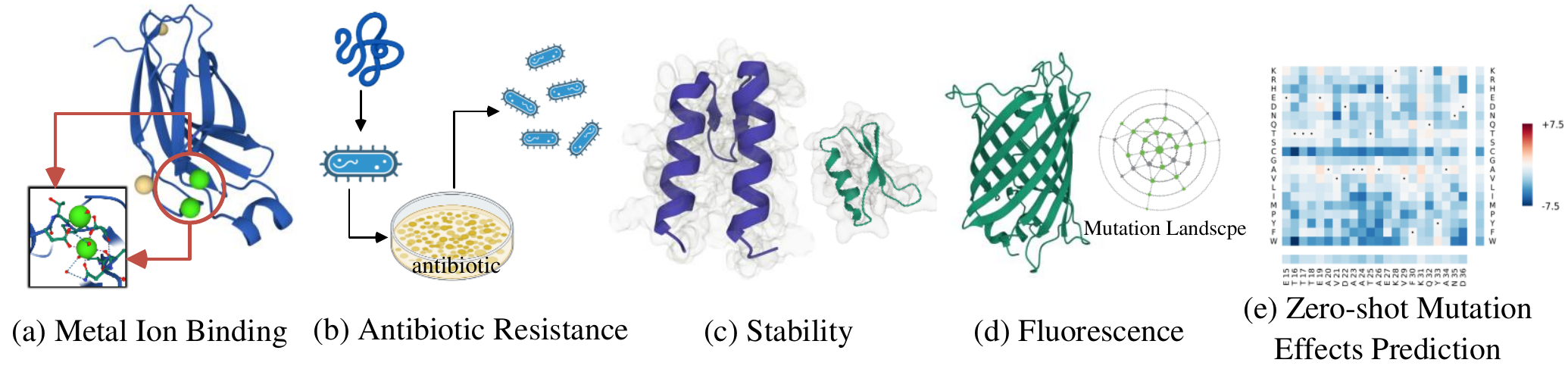}
  \caption{Protein function prediction tasks.
  (a) and (b) are annotation prediction tasks. (c)(d) and (e) are fitness prediction tasks.}
  \label{func_fig}
\end{figure}

While remarkable progress has been made in structure prediction, it is unknown whether this AlphaFold-triggered revolution transfers to other tasks, in particular function prediction. Protein function is a broad term that refers to any biological or biochemical roles in organisms.
In this paper, we focus on functional annotation prediction (classification task)~\cite{zhou2019cafa,bileschi2022using} and fitness prediction (including both regression ~\cite{dallago2021flip,rao2019evaluating} and zero-shot prediction~\cite{hie2021learning,meier2021language} tasks), as shown in Figure~\ref{func_fig}. Compared with annotation prediction, fitness prediction is more challenging since (1) there are routinely few or no laboratory labels for supervision; (2) protein AA sequences are highly similar given the same wild-type protein.

\section{Preliminaries}
\subsection{Tasks and datasets}
\label{task-des}


We evaluate models on two structure prediction tasks:
\begin{enumerate}
    \item Secondary structure (SS): This is a residue-level sequence-to-sequence task where each residue $x_i$ of a protein sequence $x = \{x_1, x_2,...,x_L\}$ is mapped to a label $y_i$ corresponding to one of eight secondary structure tasks $y_i \in$ $\{ G, H, ...,C\}$~\cite{rao2019evaluating}. SS prediction examines the degree to which a PLM learns local structure. 
    \item Contacts: For a given protein structure, two residues are considered to be in contact if their C$_{\beta}$
    carbons are within 8\AA. We evaluate on pairs that are more than 6 
    positions apart in the primary structure~\cite{rao2019evaluating}. We measure the results using Precision@$L$, which stands for the precision for the top-$L$ pairs with the highest predicted contact probability~\cite{zhang2021co}. $L$ is the length of the protein sequence.
\end{enumerate}

For both contacts and secondary structure, we use the dataset in~\cite{rives2021biological} which is constructed from SCOPe~\cite{scope}, and use the suggested split as the training and testing sets (see Table~\ref{Dataset_sizes}).
One concern is that the dataset used here has been trained by AlphaFold as they come
from the Protein Data Bank (PDB)~\cite{bank1971protein}. Hence, we investigate 48 additional proteins, which were collected from CAMEO\footnote{https://www.cameo3d.org/} (Continuous Automated Model EvaluatiOn) with `hard' category from 2021-08-28 to 2022-04-30. 


We also evaluate on two function (annotation) classification tasks:

\begin{enumerate}
\item Metal ion binding (MIB): This is a binary classification task, where a PLM with a new classification layer is used to determine whether there are metal ion–binding sites in the protein. The dataset is also collected from PDB  with annotation as metal ion binding. We randomly sample the same amount of proteins from the database as the negative class.
\item Antibiotic resistance (ABR): This is a multi-class classification task, where a PLM need to correctly determine which class of antibiotic a protein degrades. We construct the dataset from CARD~\cite{mcarthur2013comprehensive} which contains 19 different antibiotic types (see Appendix~\ref{antibioticresistance} for details).
\end{enumerate}

And three fitness prediction tasks. Unlike functional annotation prediction, protein sequences in this task are all from the same wild-type with a small number of mutated residues. 

\begin{enumerate}
\item Stability: This is a protein-level regression task that predicts the protease concentration at which a protein can maintain its fold~\cite{rocklin2017global}. We use the data splits from TAPE. 
\item Fluorescence: This is also a protein-level regression task, predicting the log-fluorescence intensity of the protein sequence~\cite{sarkisyan2016local}. We use the data splits from TAPE. 
\item Zero-shot mutation effect prediction: This is a protein-level prediction task by comparing the difference between likelihoods assigned to the
mutated residue and the likelihoods assigned to the wild-type (see~\cite{meier2021language} for details). We evaluate five protein mutation datasets from DeepSequence~\cite{riesselman2018deep}. Only single point mutation data is considered in this sub-task.
\end{enumerate}


\subsection{Methods} 
For supervised function prediction tasks, we remove the original classification layer for ESM-1b and MSA-Transformer, and the Structure module for AlphaFold. These remaining parts are called protein representations. We then add a linear layer on top of these representation models to perform new classification or regression tasks. We adopt the standard fine-tuning strategy by fine-tuning all parameters using Adamw optimizer with 1e-5 as learning rate.
For the zero-shot fitness prediction task, no training is needed. Instead, we obtain the softmax output at the corresponding mutation site in each protein sequence. The probability value is regarded as the fitness value for each amino acid following~\cite{hie2021learning,meier2021language}. As for structure predictions, we follow the same practice in~\cite{rao2019evaluating}. Other details can be seen in Appendix~\ref{methoddetails}.

\begin{table}
  \caption{Dataset descriptions}
  \label{Dataset_sizes}
  \centering
  \begin{tabular}{lrrr}
    \toprule
    Task     & Source & Train     & Test \\
    \midrule
    Secondary Structure \& Contact Prediction & SCOPe & 11680 & 3617 \\
    Metal Ion Binding & PDB & 6000 & 1332 \\
    Antibiotic Resistance & CARD & 2072 & 530 \\
    Fluorescence & TAPE & 21446 & 27217 \\
    Stability & TAPE & 53679 & 12839 \\
    \bottomrule
  \end{tabular}
\end{table}

\section{Results}
\subsection{Structure Prediction}\label{section_Structure_Predition}
In this section, we examine the structure representation ability of the three PLMs models discussed above. In general, the secondary structure and contact prediction tasks do not have significant practical values since there is already highly accurate 3D structure data by AlphaFold. The purpose instead is to provide a reference for function prediction, given that some tasks have similar formulation.

Table~\ref{SSP_res} and ~\ref{cm_res} (Appendix~\ref{sec:contact}) show the results of SS prediction and contact prediction, respectively. For SS prediction, we consider two settings: pre-trained parameters and training from scratch (i.e., with random initialization). For  Evoformer, pre-training means the training of AlphaFold. 
For contact prediction, all PLMs have been already pre-trained since there is no meaningful contact map without pre-training.

\begin{figure}
  \centering
  \includegraphics[width = 1.0\textwidth]{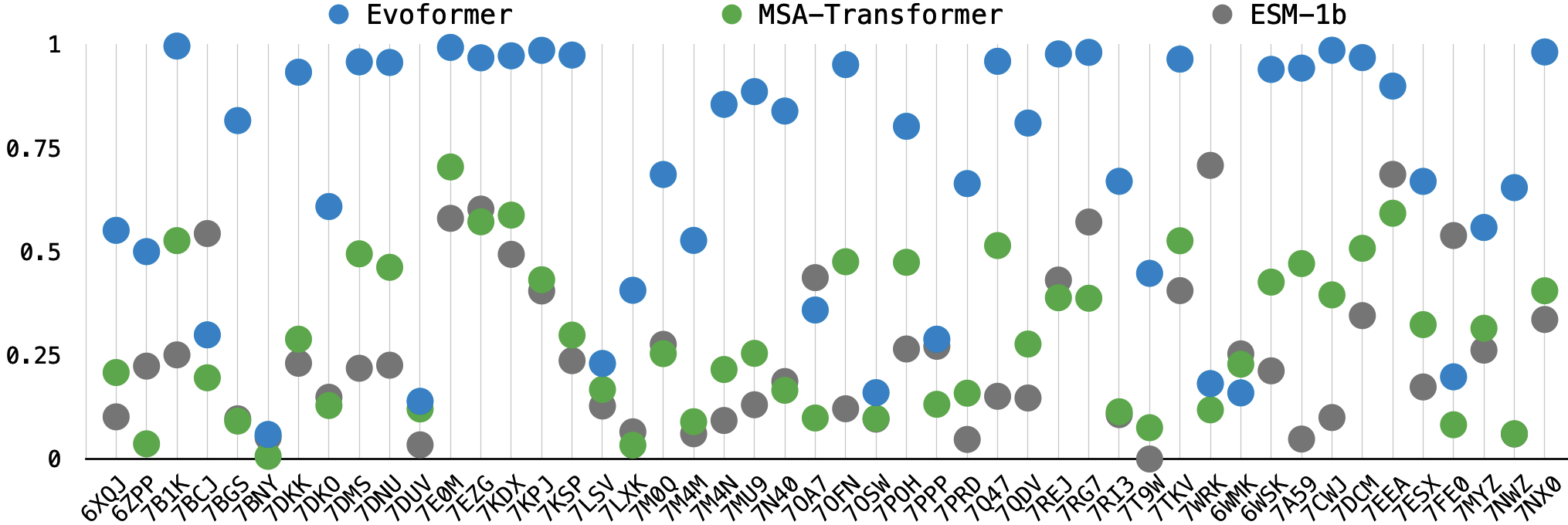}
  \caption{ Contact map prediction with the CAMEO dataset (from 2021-08-28 to 2022-04-30) in terms of Precision@$L$. These proteins have not been used for training AlphaFold.}
  \label{contact_map_fig}
\end{figure}

First, we can easily observe that with pre-training, Evoformer performs the best  in both tasks. Particularly, Evoformer outperforms ESM-1b and MSA-Transformer by a large margin for contact prediction with over 94\% accuracy on Precision@$L$ (see Table~\ref{cm_res}). 
By contrast, it does not perform as accurately as the contact prediction task for SS prediction, which improves ESM-1b by 11.6\% and MSA-Transformer by 5\%. This is reasonable since SS prediction is  different from 3D structure prediction from the machine learning perspective. Even though Evoformer has strong structure representation ability, it may not work well by just adding a linear classification layer. This suggests that a more complex structure module for SS prediction is necessary for higher accuracy.
In comparison, since contact information can be directly extracted from the pairwise distance map, it is not surprising that Evoformer shows superb results in this task. Thereby, we have also demonstrated results of 48 new proteins in Figure~\ref{contact_map_fig}, where these proteins have not been trained by AlphaFold. In general, they are consistent with our above  analysis.

Second, we observe that pre-training is important: all PLMs are remarkably improved with around 40\% improvement for ESM-1b and 28\% for Evoformer. Interestingly, we note that  Evoformer performs worse than MSA-Transformer by training from scratch. We conjecture that Evoformer is much harder to be trained without good initialization as its network architecture is more complex and deeper than MSA-Transformer. 

Finally, for structure prediction, evolution-aware models are largely better than  evolution-free ESM-1b. This is consistent with biological intuition and  previous observation, such as in~\cite{rao2021msa}.

\begin{table}
  \caption{SS prediction. `Scratch' means training from scratch without pre-training.}
  \label{SSP_res}
  \centering
  \begin{tabular}{lrrrr}
    \toprule
    Model          & Pre-train &Scratch &Improv. \\
    \midrule
    ESM-1b            & 0.703& 0.500 &40.6\%  \\
    MSA-Transformer         & 0.748 & 0.634 &18.0\% \\
        Evoformer           & \textbf{0.785} & 0.614 &27.8\% \\
    \bottomrule
  \end{tabular}
\end{table}

\subsection{Supervised Function Prediction}
\label{superfunction_pred}
AlphaFold predicts very accurate protein structures. However,
it remains unknown whether its core representation module Evoformer can prediction function. Similarly, we investigate MSA-Transformer's ability to predict protein function, \textcolor{black}{which is unknown either}. 
Note that, unlike annotation prediction, evolution-aware Evoformer and MSA-Transformer only take single sequences (rather than MSAs) as input since all protein variants share the same MSAs for the fitness prediction task.

The results of annotation prediction (protein-level classification) and fitness predictions  (protein-level regression) are listed in Table~\ref{function_prediction} and ~\ref{fitness_prediction} respectively. We use classification accuracy~\cite{rao2019evaluating} for evaluating annotation prediction and Spearman rank correlation~\cite{meier2021language} for evaluating fitness prediction.
First,
we find that for annotation prediction, ESM-1b, the evolution-free PLM, exceeds the two evolution-aware PLMs, although opposite observations are made for predicting structures. Specifically, pre-trained ESM-1b achieves 5.8\% and 17.4\% improvements over pre-trained Evoformer and MSA-Transformer. This is a bit surprising since the biological intuition is that protein functional properties are mediated by structures. An important conclusion we reached here is that \textbf{better structure PLMs do not mean they have a better representation  for predicting function.} 

Second, all pre-trained PLMs perform very well on the ABR task, a multi-class protein classification task. Despite that, we can still observe an obvious improvement between training-from-scratch and pre-training. Similar observation brought from pre-training can be obviously seen from all other function prediction tasks. This suggests that \textbf{both supervised and unsupervised pre-training on PLMs are very useful to obtain protein function representations.}

Third, we observe that the three pre-trained  PLMs  in general show better results than the one-hot~\cite{rao2019evaluating} and ResNet~\cite{rao2019evaluating,he2016deep} baselines in Table~\ref{fitness_prediction}. This observation is new since neither Evoformer nor MSA-Transformer have been investigated for supervised fitness prediction.
In more detail, ESM-1b performs the best on the fluorescence task, whereas Evoformer is the best on the stability task. This may be because protein stability has a closer relationship with protein structure, or because sequence-only pretraining does not generalize as well as structural pretraining to \textit{de novo} miniproteins. 


\begin{table}
  \caption{Functional annotation prediction. 'scratch' means random initialization for parameters.}
  \label{function_prediction}
  \centering
  \begin{tabular}{lrrrr}
    \toprule
            &\multicolumn{2}{c}{Pre-train} & \multicolumn{2}{c}{Scratch}\\
    \cmidrule(lr){2-3}\cmidrule(lr){4-5}
    Model   & MIB&ABR&MIB&ABR\\
    \midrule
    ESM-1b  & \textbf{0.840} & \textbf{0.979} & 0.628 & 0.945\\
    MSA-Transformer & 0.715 & 0.961 & 0.640 & 0.932\\
       Evoformer  & 0.794 & 0.979 & 0.645 & 0.920\\
    \bottomrule
  \end{tabular}
\end{table}

\begin{table}
  \caption{Fitness prediction. Scores are |Spearman $\rho$| on each task.}
  \label{fitness_prediction}
  \centering
  \begin{tabular}{lrrrr}
    \toprule
    &\multicolumn{2}{c}{Pre-train} & \multicolumn{2}{c}{Scratch}\\
    \cmidrule(lr){2-3}\cmidrule(lr){4-5}
    Model&Fluorescence&Stability&Fluorescence&Stability\\
    \midrule
    One-hot~\cite{rao2019evaluating}&0.14&0.19&-&-\\
    ResNet~\cite{rao2019evaluating}&0.21&0.73&0.28&0.61\\
    ESM-1b&\textbf{0.68}&0.76&0.68&0.59\\
    MSA-Transformer&0.64&0.67&0.67&0.61\\
    Evoformer&0.67&\textbf{0.79}&0.36&0.52\\
    \bottomrule
  \end{tabular}
\end{table}

\subsection{Zero-shot Mutation effects Prediction}
 Recent work revealed that PLMs are strong zero-shot learners in predicting potential viral mutations~\cite{hie2021learning,hie2022evolocity}. We are interested in evaluating Evoformer's masked-residue reconstruction ability in such a zero-shot setting given that it also has a BERT-like masked token prediction loss.

\begin{figure}
  \centering
  \includegraphics[width = 1\textwidth]{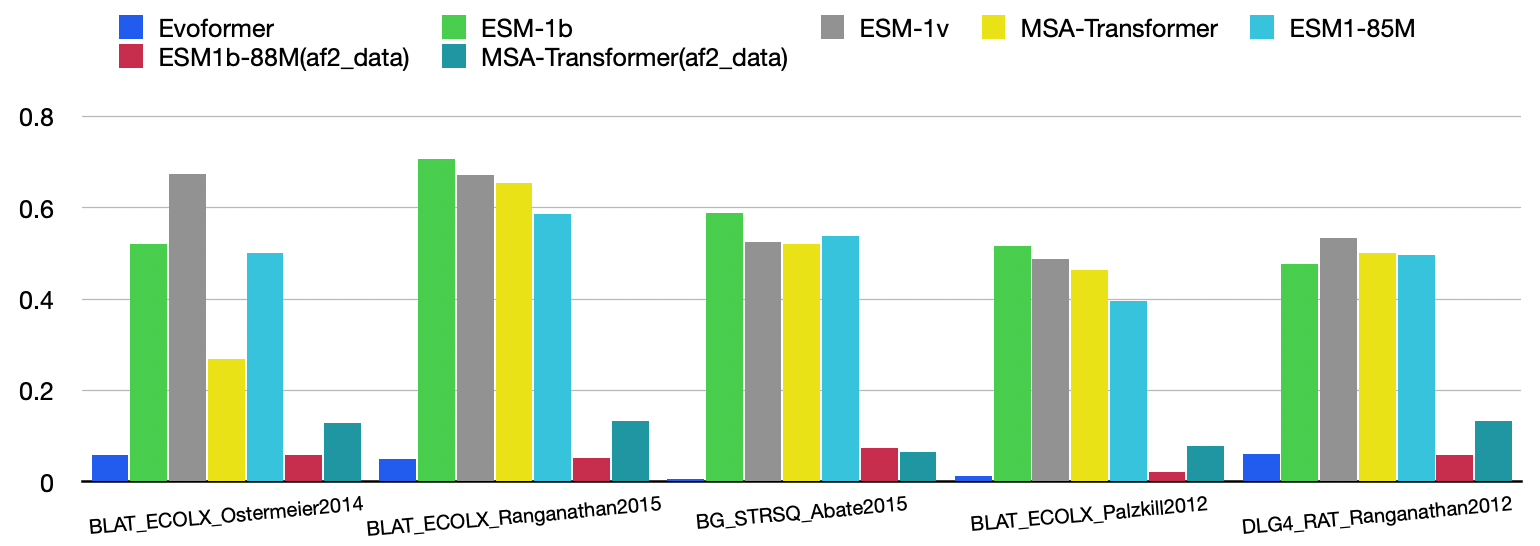}
  \caption{Zero-shot fitness prediction. ESM-1v's results is the average performance of five models; ESM1-85M  was trained with the same dataset as ESM-1b but with a much smaller model size (85 million parameters); ESM1b-88M(af2\_data) and MSA-Transformer(af2\_data) were trained with the same protein dataset as used for training AlphaFold  
  (including both the PDB data and those from Uniclust30 used in the self-distillation process) and have similar model sizes as Evoformer.} 
  \label{mutation}
\end{figure}

Figure~\ref{mutation} shows the results of all PLMs. First, we observe that ESM-1b and ESM-1v~\cite{meier2021language} yield the best results, and both of them are evolution-free models. ESM-1v shares an architecture with ESM-1b, but is an ensemble of 5 models trained on UniRef90 instead of UniRef50. 
MSA-Transformer in general performs well except on BLAT\_ECOLX\_Ostermeier2014. We notice that MSA-Transformer is highly affected by the quality and  sequence identity thresholds of MSAs.
For our experiments, we use the default setting by HHblits~\cite{remmert2012hhblits} searching from BFD\footnote{https://bfd.mmseqs.com/} database (see Appendix~\ref{hhblitsetting}).


The most surprising results are for Evoformer, which performs poorly in this task on all five datasets. 
To examine this cause, we present results of another three models, namely ESM1-85M (released by ~\cite{rives2021biological}), ESM1b-88M (af2\_data) and MSA-Transformer (af2\_data)
(see model descriptions in Figure~\ref{mutation}). 
As shown, ESM1-85M performs much better than ESM1b-88M, which indicates that the accuracy of zero-shot fitness prediction task is largely affected by the size of the pre-training dataset. This may explain why Evoformer performs so poorly in this task. Beyond this, there is another key difference between these models. Evoformer or AlphaFold were trained with five different loss functions, among which four are structure-related loss and
 only one auxiliary loss is the masked token reconstruction (MTR) loss. By contrast, MSA-Transformer and ESM-1b were trained with only the MTR loss. This may also be a reason for explaining the worse results of Evoformer.

\textbf{Answer for Q(\romannumeral1):} \textbf{Yes}. The answer is provided in Table~\ref{function_prediction} and ~\ref{fitness_prediction}. From these results, we can see that  Evoformer with pre-trained parameters largely outperforms the non-pretrained version in four different function prediction tasks. Together with
its structure representation ability, we conclude that \textbf{parameters learned by AlphaFold are general-purpose and useful to various  structure and function prediction tasks.} 

\textbf{Answer for Q(\romannumeral2):} \textbf{No}. The answer is provided in Table~\ref{SSP_res},~\ref{cm_res},~\ref{function_prediction},~\ref{fitness_prediction} and Figure~\ref{contact_map_fig}.  These results show that Evoformer indeed outperforms MSA-Transformer and ESM-1b when predicting structures, but it does not obviously outperform ESM-1b when predicting function. In particular,
it does not work at all in the zero-shot mutation effect prediction task.
Accordingly, we can conclude that \textbf{the AlphaFold-triggered revolution for structure prediction cannot be directly transferred to function predictions. ESM-1b is still the SOTA when predicting protein functions.}

\subsection{Effect of MSA}
Here, we want to understand the influence of MSAs on evolution-aware PLMs.  Further, we 
 explore the relationship between evolution-aware  and -free models.
 As mentioned in Section~\ref{superfunction_pred}, evolution-aware PLMs  take only a single sequence as input for the supervised fitness prediction tasks, so we do not consider the affect of MSAs for these two tasks. 
\begin{table}
  \caption{Impact of MSAs. 'Seq' denotes an individual sequence, i.e., no MSAs.}
  \label{num_msa_table}
  \centering
  \begin{tabular}{lrrrrrrr}
    \toprule
        &   &\multicolumn{2}{c}{SS} & \multicolumn{2}{c}{MIB} & \multicolumn{2}{c}{ABR}\\
    \cmidrule(lr){3-4}\cmidrule(lr){5-6}\cmidrule(lr){7-8}
    Model&Pretrained&MSA&Seq&MSA&Seq&MSA&Seq\\
    \midrule
    Evoformer&Yes&0.785&0.716&0.794&0.724 & 0.979&0.983 \\
    MSA-Transformer&Yes&0.748&0.631&0.715&0.707  & 0.961&0.908\\
    Evoformer&No&0.614&0.624&0.645&0.632   & 0.920&0.875\\ 
    MSA-Transformer&No&0.634&0.526&0.640&0.579   & 0.932&0.909\\
    \bottomrule
  \end{tabular}
\end{table}

\begin{figure}
  \centering
  \includegraphics[width = 0.87\textwidth]{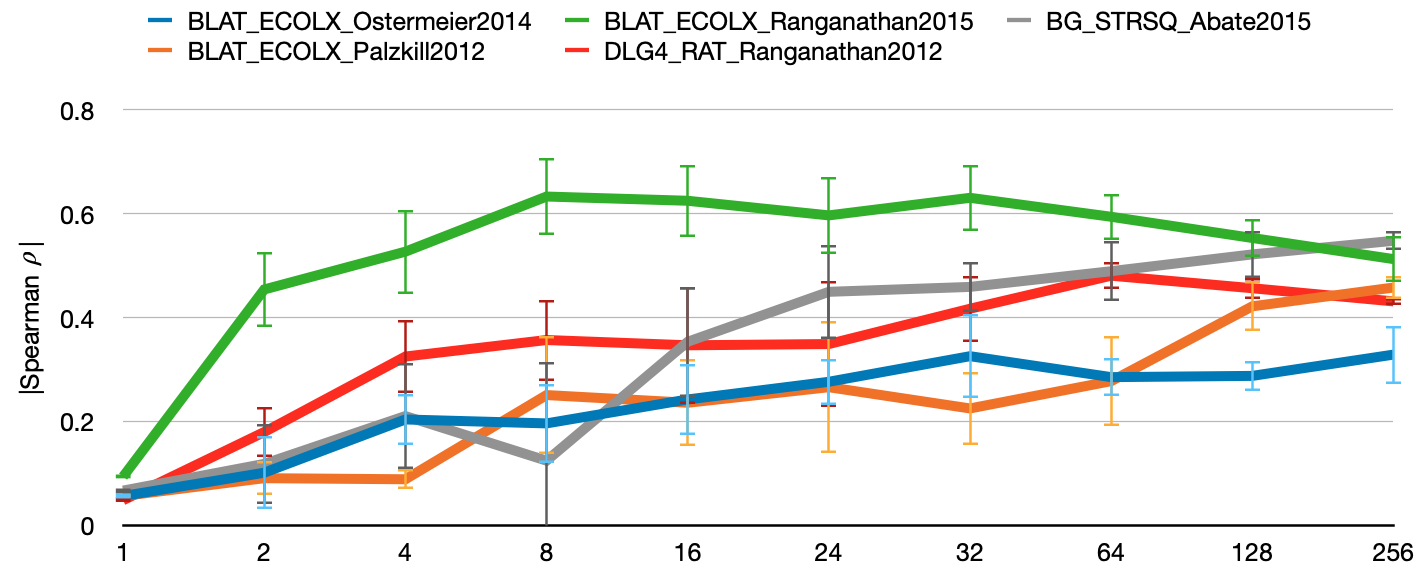}
  \caption{The effect of  MSA depth on the zero-shot fitness prediction task using MSA-Transformer. Error bars is the  standard deviation of six replicate  experiments with randomly sampled $k$ MSAs. }
  \label{num_msa_graph}
\end{figure}

We report results on Table~\ref{num_msa_table} and Figure~\ref{num_msa_graph}. The MSA setting in  Table~\ref{num_msa_table}  is consistent for both model training and inference. For Figure~\ref{num_msa_graph}, we only consider the inference stage because it is a zero-shot task.
Clearly, we observe that without MSAs, Evoformer, and MSA-Transformer both yield much worse results (with or without pre-training). Similarly, as shown in Figure~\ref{num_msa_graph}, MSA-Transformer is also affected by the number of MSAs.
In general, it performs much worse with few MSAs.
These results show that (\textbf{Answer for Q(\romannumeral3)}) \textbf{MSA data is a crucial input for evolution-aware PLMs, for not only the structure prediction but also the function prediction tasks.}


\subsection{Relationship between Evolution-aware \& -free PLMs}
Evolution-aware PLMs require MSA data as input for both training and inference. However, database search using HHblits~\cite{remmert2012hhblits} or Jackhmmer~\cite{johnson2010hidden} can take up to 10 minutes for a 1000-residue protein.  
On the other hand, given the powerful function representation of ESM-1b, we design a fast homology retrieval framework using Siamese ESM-1b, following~\cite{reimers2019sentence}. Figure~\ref{esm1b_MSA} is the schematic of the ESM-1b-constructed MSA generating and serving process, called ESM-MSA (see Appendix~\ref{ESM-MSAdetails}). 
\begin{table}[H]
  \caption{MSA Impacts. `Seq' denotes an individual sequence i.e. no MSAs.}
  \label{generate_table}
  \centering
  \begin{tabular}{ccccccc}
    \toprule
       &\multicolumn{3}{c}{SS Prediction} & \multicolumn{3}{c}{MIB}\\
    \cmidrule(lr){2-4}\cmidrule(lr){5-7}
    Model  & Jackhmmer & ESM-MSA & Seq & Jackhmmer & ESM-MSA & Seq\\
    \midrule
    Evoformer &0.785 &0.776 &0.716  & 0.794& 0.766  &0.724       \\
    MSA-Transformer& 0.748&0.733& 0.631 &0.715 &0.774 &0.707     \\
    \bottomrule
  \end{tabular}
\end{table}
Table \ref{generate_table} shows the results by using different MSAs for Evoformer and MSA-Transformer. MSAs generated by ESM-1b are comparable to those found by Jackhmmer (or HHblits)  while being much faster (see Appendix~\ref{speedmsa}), 
suggesting that (\textbf{Answer for Q(\romannumeral3)}) \textbf{the function representation learned by evolution-free ESM-1b can be used to construct MSAs for evolution-aware MSA-Transformer and Evoformer.}




\section{Conclusions and Limitations}
In the paper, we have presented empirical studies to characterize three  successful PLMs. In particular, we focus on the function representation capacity of Evoformer. We draw several important conclusions: (i) Evoformer encodes not only structure, but also various protein functional properties; it provides an alternative choice when predicting protein functions, especially the stability prediction task; (ii) despite that
AlphaFold dominates the structure prediction tasks, Evoformer is not yet able to substitute ESM-1b and MSA-Transformer when predicting protein functions; 
(iii) the functional predictability of evolution-aware PLMs is also influenced by protein MSAs. Interestingly, evolution-free ESM-1b can generate  sufficiently-accurate MSAs for evolution-aware PLMs with  much higher efficiency. 
Our work points out both strengths  and weakness of three SOTA PLMs, which are potentially useful to many downstream biological tasks, including but not limited to protein engineering~\cite{biswas2021low},  cancer early detection~\cite{frazer2021disease}, and drug  discovery~\cite{kalakoti2022transdti}.

An important limitation is that we probe the representation capacity of these PLMs by only adding a linear head on top --- a common practice in the deep learning community. While this verifies our findings, we believe such a linear layer is not sufficiently expressive to achieve SOTA results. That is the reason why Evoformer with a eight-layer Structure module achieves much higher 3D structure accuracy but with a linear layer it achieves less than 80\% classification accuracy for the secondary structure prediction task. In addition, we do not fully disentangle the effects of the Evoformer architecture, training data, and loss function when comparing it to MSA-Transformer. Therefore, an important direction is to develop more advanced functional modules for higher prediction accuracy.
\section*{Acknowledgement}
This work is supported by special funding from the Westlake Center of Synthetic Biology and Integrated Bioengineering (WE-SynBio),
the National Natural Science Foundation of China (No. U21A20427), and the Research Center for Industries of the Future (No. WU2022C030).
\newpage
\printbibliography

@article{jumper2021highly,
  title={Highly accurate protein structure prediction with AlphaFold},
  author={Jumper, John and Evans, Richard and Pritzel, Alexander and Green, Tim and Figurnov, Michael and Ronneberger, Olaf and Tunyasuvunakool, Kathryn and Bates, Russ and {\v{Z}}{\'\i}dek, Augustin and Potapenko, Anna and others},
  journal={Nature},
  volume={596},
  number={7873},
  pages={583--589},
  year={2021},
  publisher={Nature Publishing Group}
}

@article{teufel2021signalp,
  title={Signal{P} 6.0 achieves signal peptide prediction across all types using protein language models},
  author={Teufel, Felix and Armenteros, Jos{\'e} Juan Almagro and Johansen, Alexander Rosenberg and G{\'\i}slason, Magn{\'u}s Halld{\'o}r and Pihl, Silas Irby and Tsirigos, Konstantinos D and Winther, Ole and Brunak, S{\o}ren and von Heijne, Gunnar and Nielsen, Henrik},
  journal={bioRxiv},
  year={2021},
  publisher={Cold Spring Harbor Laboratory}
}

@article{heinzinger2019modeling,
  title={Modeling aspects of the language of life through transfer-learning protein sequences},
  author={Heinzinger, Michael and Elnaggar, Ahmed and Wang, Yu and Dallago, Christian and Nechaev, Dmitrii and Matthes, Florian and Rost, Burkhard},
  journal={BMC bioinformatics},
  volume={20},
  number={1},
  pages={1--17},
  year={2019},
  publisher={BioMed Central}
}

@article{bileschi2019annotate,
  title        = {Using Deep Learning to Annotate the Protein Universe},
  author       = {Bileschi, Maxwell L. and Belanger, David and Bryant, Drew and Sanderson, Theo and Carter, Brandon and Sculley, D. and DePristo, Mark A. and Colwell, Lucy J.},
  year         = 2019,
  journal      = {bioRxiv},
  publisher    = {Cold Spring Harbor Laboratory},
  doi          = {10.1101/626507},
  url          = {https://www.biorxiv.org/content/early/2019/05/06/626507},
  elocation-id = 626507,
  eprint       = {https://www.biorxiv.org/content/early/2019/05/06/626507.full.pdf}
}

@article{hie2022evolocity,
   author = {Brian L. Hie and Kevin K. Yang and Peter S. Kim},
   doi = {10.1016/j.cels.2022.01.003},
   issn = {24054712},
   journal = {Cell Systems},
   month = {2},
   title = {Evolutionary velocity with protein language models predicts evolutionary dynamics of diverse proteins},
   year = {2022},
}

@article{riesselman2018deep,
  title        = {Deep generative models of genetic variation capture the effects of mutations},
  author       = {Riesselman, Adam J and Ingraham, John B and Marks, Debora S},
  year         = 2018,
  journal      = {Nature Methods},
  publisher    = {Nature Publishing Group},
  volume       = 15,
  number       = 10,
  pages        = {816--822}
}

@article{thumuluri2022deeploc,
  title={Deep{L}oc 2.0: multi-label subcellular localization prediction using protein language models},
  author={Thumuluri, Vineet and Almagro Armenteros, Jos{\'e} Juan and Johansen, Alexander Rosenberg and Nielsen, Henrik and Winther, Ole},
  journal={Nucleic Acids Research},
  year={2022}
}

@inproceedings{rao2021msa,
  title={MSA transformer},
  author={Rao, Roshan M and Liu, Jason and Verkuil, Robert and Meier, Joshua and Canny, John and Abbeel, Pieter and Sercu, Tom and Rives, Alexander},
  booktitle={International Conference on Machine Learning},
  pages={8844--8856},
  year={2021},
  organization={PMLR}
}

@article{rives2021biological,
  title={Biological structure and function emerge from scaling unsupervised learning to 250 million protein sequences},
  author={Rives, Alexander and Meier, Joshua and Sercu, Tom and Goyal, Siddharth and Lin, Zeming and Liu, Jason and Guo, Demi and Ott, Myle and Zitnick, C Lawrence and Ma, Jerry and others},
  journal={Proceedings of the National Academy of Sciences},
  volume={118},
  number={15},
  year={2021},
  publisher={National Acad Sciences}
}

@article{scope,
    author = {Fox, Naomi K. and Brenner, Steven E. and Chandonia, John-Marc},
    title = "{SCOPe: Structural Classification of Proteins—extended, integrating SCOP and ASTRAL data and classification of new structures}",
    journal = {Nucleic Acids Research},
    volume = {42},
    number = {D1},
    pages = {D304-D309},
    year = {2013},
    month = {12},
    issn = {0305-1048},
    doi = {10.1093/nar/gkt1240},
    eprint = {https://academic.oup.com/nar/article-pdf/42/D1/D304/3647135/gkt1240.pdf},
}

@article{rao2019evaluating,
  title={Evaluating protein transfer learning with TAPE},
  author={Rao, Roshan and Bhattacharya, Nicholas and Thomas, Neil and Duan, Yan and Chen, Peter and Canny, John and Abbeel, Pieter and Song, Yun},
  journal={Advances in neural information processing systems},
  volume={32},
  year={2019}
}

@article{elnaggar2020prottrans,
  title={ProtTrans: towards cracking the language of Life's code through self-supervised deep learning and high performance computing},
  author={Elnaggar, Ahmed and Heinzinger, Michael and Dallago, Christian and Rihawi, Ghalia and Wang, Yu and Jones, Llion and Gibbs, Tom and Feher, Tamas and Angerer, Christoph and Steinegger, Martin and others},
  journal={arXiv preprint arXiv:2007.06225},
  year={2020}
}

@article{bepler2021learning,
  title={Learning the protein language: Evolution, structure, and function},
  author={Bepler, Tristan and Berger, Bonnie},
  journal={Cell systems},
  volume={12},
  number={6},
  pages={654--669},
  year={2021},
  publisher={Elsevier}
}

@article{HHblits,
   author = {Remmert, Michael and Biegert, Andreas and Hauser, Andreas and Söding, Johannes},
   title = {HHblits: lightning-fast iterative protein sequence searching by HMM-HMM alignment},
   journal = {Nature Methods},
   volume = {9},
   number = {2},
   pages = {173-175},
   ISSN = {1548-7105},
   DOI = {10.1038/nmeth.1818},
   year = {2012},
   type = {Journal Article}
}

@article{devlin2018bert,
  title={Bert: Pre-training of deep bidirectional transformers for language understanding},
  author={Devlin, Jacob and Chang, Ming-Wei and Lee, Kenton and Toutanova, Kristina},
  journal={arXiv preprint arXiv:1810.04805},
  year={2018}
}

@article{alley2019unified,
  title={Unified rational protein engineering with sequence-based deep representation learning},
  author={Alley, Ethan C and Khimulya, Grigory and Biswas, Surojit and AlQuraishi, Mohammed and Church, George M},
  journal={Nature methods},
  volume={16},
  number={12},
  pages={1315--1322},
  year={2019},
  publisher={Nature Publishing Group}
}

@article{bepler2019learning,
  title={Learning protein sequence embeddings using information from structure},
  author={Bepler, Tristan and Berger, Bonnie},
  booktitle={International Conference on Learning Representations},
  year={2019}
}

@article{vaswani2017attention,
  title={Attention is all you need},
  author={Vaswani, Ashish and Shazeer, Noam and Parmar, Niki and Uszkoreit, Jakob and Jones, Llion and Gomez, Aidan N and Kaiser, {\L}ukasz and Polosukhin, Illia},
  journal={Advances in neural information processing systems},
  volume={30},
  year={2017}
}

@article{sussman1998protein,
  title={Protein Data Bank (PDB): database of three-dimensional structural information of biological macromolecules},
  author={Sussman, Joel L and Lin, Dawei and Jiang, Jiansheng and Manning, Nancy O and Prilusky, Jaime and Ritter, Otto and Abola, Enrique E},
  journal={Acta Crystallographica Section D: Biological Crystallography},
  volume={54},
  number={6},
  pages={1078--1084},
  year={1998},
  publisher={International Union of Crystallography}
}

@article{zhou2019cafa,
  title={The CAFA challenge reports improved protein function prediction and new functional annotations for hundreds of genes through experimental screens},
  author={Zhou, Naihui and Jiang, Yuxiang and Bergquist, Timothy R and Lee, Alexandra J and Kacsoh, Balint Z and Crocker, Alex W and Lewis, Kimberley A and Georghiou, George and Nguyen, Huy N and Hamid, Md Nafiz and others},
  journal={Genome biology},
  volume={20},
  number={1},
  pages={1--23},
  year={2019},
  publisher={BioMed Central}
}

@article{dallago2021flip,
  title={FLIP: Benchmark tasks in fitness landscape inference for proteins},
  author={Dallago, Christian and Mou, Jody and Johnston, Kadina E and Wittmann, Bruce J and Bhattacharya, Nicholas and Goldman, Samuel and Madani, Ali and Yang, Kevin K},
    journal={Advances in Neural Information Processing Systems},
  year={2021}
}

@article{meier2021language,
  title={Language models enable zero-shot prediction of the effects of mutations on protein function},
  author={Meier, Joshua and Rao, Roshan and Verkuil, Robert and Liu, Jason and Sercu, Tom and Rives, Alex},
  journal={Advances in Neural Information Processing Systems},
  volume={34},
  year={2021}
}

@article{evans2021protein,
  title={Protein complex prediction with AlphaFold-Multimer},
  author={Evans, Richard and O'Neill, Michael and Pritzel, Alexander and Antropova, Natasha and Senior, Andrew W and Green, Timothy and {\v{Z}}{\'\i}dek, Augustin and Bates, Russell and Blackwell, Sam and Yim, Jason and others},
  journal={BioRxiv},
  year={2021},
  publisher={Cold Spring Harbor Laboratory}
}

@inproceedings{rao2020transformer,
  title={Transformer protein language models are unsupervised structure learners},
  author={Rao, Roshan and Meier, Joshua and Sercu, Tom and Ovchinnikov, Sergey and Rives, Alexander},
  booktitle={International Conference on Learning Representations},
  year={2020}
}

@article{baek2021accurate,
  title={Accurate prediction of protein structures and interactions using a three-track neural network},
  author={Baek, Minkyung and DiMaio, Frank and Anishchenko, Ivan and Dauparas, Justas and Ovchinnikov, Sergey and Lee, Gyu Rie and Wang, Jue and Cong, Qian and Kinch, Lisa N and Schaeffer, R Dustin and others},
  journal={Science},
  volume={373},
  number={6557},
  pages={871--876},
  year={2021},
  publisher={American Association for the Advancement of Science}
}

@article{sonderby2014protein,
  title={Protein secondary structure prediction with long short term memory networks},
  author={S{\o}nderby, S{\o}ren Kaae and Winther, Ole},
  journal={arXiv preprint arXiv:1412.7828},
  year={2014}
}

@article{wang2017accurate,
  title={Accurate de novo prediction of protein contact map by ultra-deep learning model},
  author={Wang, Sheng and Sun, Siqi and Li, Zhen and Zhang, Renyu and Xu, Jinbo},
  journal={PLoS computational biology},
  volume={13},
  number={1},
  pages={e1005324},
  year={2017},
  publisher={Public Library of Science San Francisco, CA USA}
}

@article{hie2021learning,
  title={Learning the language of viral evolution and escape},
  author={Hie, Brian and Zhong, Ellen D and Berger, Bonnie and Bryson, Bryan},
  journal={Science},
  volume={371},
  number={6526},
  pages={284--288},
  year={2021},
  publisher={American Association for the Advancement of Science}
}

@article{biswas2021low,
  title={Low-N protein engineering with data-efficient deep learning},
  author={Biswas, Surojit and Khimulya, Grigory and Alley, Ethan C and Esvelt, Kevin M and Church, George M},
  journal={Nature methods},
  volume={18},
  number={4},
  pages={389--396},
  year={2021},
  publisher={Nature Publishing Group}
}

@article{bileschi2022using,
  title={Using deep learning to annotate the protein universe},
  author={Bileschi, Maxwell L and Belanger, David and Bryant, Drew H and Sanderson, Theo and Carter, Brandon and Sculley, D and Bateman, Alex and DePristo, Mark A and Colwell, Lucy J},
  journal={Nature Biotechnology},
  pages={1--6},
  year={2022},
  publisher={Nature Publishing Group}
}

@article{UniProt,
   author = {UniProt, Consortium},
   title = {The universal protein resource (UniProt)},
   journal = {Nucleic acids research},
   volume = {36},
   number = {Database issue},
   pages = {D190-D195},
   ISSN = {1362-4962
0305-1048},
   year = {2008},
   type = {Journal Article}
}

@article{vig2020bertology,
  title={Bertology meets biology: Interpreting attention in protein language models},
  author={Vig, Jesse and Madani, Ali and Varshney, Lav R and Xiong, Caiming and Socher, Richard and Rajani, Nazneen Fatema},
  journal={arXiv preprint arXiv:2006.15222},
  year={2020}
}

@article{sarkisyan2016local,
  title={Local fitness landscape of the green fluorescent protein},
  author={Sarkisyan, Karen S and Bolotin, Dmitry A and Meer, Margarita V and Usmanova, Dinara R and Mishin, Alexander S and Sharonov, George V and Ivankov, Dmitry N and Bozhanova, Nina G and Baranov, Mikhail S and Soylemez, Onuralp and others},
  journal={Nature},
  volume={533},
  number={7603},
  pages={397--401},
  year={2016},
  publisher={Nature Publishing Group}
}

@article{rocklin2017global,
  title={Global analysis of protein folding using massively parallel design, synthesis, and testing},
  author={Rocklin, Gabriel J and Chidyausiku, Tamuka M and Goreshnik, Inna and Ford, Alex and Houliston, Scott and Lemak, Alexander and Carter, Lauren and Ravichandran, Rashmi and Mulligan, Vikram K and Chevalier, Aaron and others},
  journal={Science},
  volume={357},
  number={6347},
  pages={168--175},
  year={2017},
  publisher={American Association for the Advancement of Science}
}

@article{zhang2021co,
  title={Co-evolution Transformer for Protein Contact Prediction},
  author={Zhang, He and Ju, Fusong and Zhu, Jianwei and He, Liang and Shao, Bin and Zheng, Nanning and Liu, Tie-Yan},
  journal={Advances in Neural Information Processing Systems},
  volume={34},
  year={2021}
}

@inproceedings{he2016deep,
  title={Deep residual learning for image recognition},
  author={He, Kaiming and Zhang, Xiangyu and Ren, Shaoqing and Sun, Jian},
  booktitle={Proceedings of the IEEE conference on computer vision and pattern recognition},
  pages={770--778},
  year={2016}
}

@article{frazer2021disease,
  title={Disease variant prediction with deep generative models of evolutionary data},
  author={Frazer, Jonathan and Notin, Pascal and Dias, Mafalda and Gomez, Aidan and Min, Joseph K and Brock, Kelly and Gal, Yarin and Marks, Debora S},
  journal={Nature},
  volume={599},
  number={7883},
  pages={91--95},
  year={2021},
  publisher={Nature Publishing Group}
}

@article{remmert2012hhblits,
  title={HHblits: lightning-fast iterative protein sequence searching by HMM-HMM alignment},
  author={Remmert, Michael and Biegert, Andreas and Hauser, Andreas and S{\"o}ding, Johannes},
  journal={Nature methods},
  volume={9},
  number={2},
  pages={173--175},
  year={2012},
  publisher={Nature Publishing Group}
}

@article{reimers2019sentence,
  title={Sentence-bert: Sentence embeddings using siamese bert-networks},
  author={Reimers, Nils and Gurevych, Iryna},
  journal={arXiv preprint arXiv:1908.10084},
  year={2019}
}

@article{kalakoti2022transdti,
  title={TransDTI: Transformer-Based Language Models for Estimating DTIs and Building a Drug Recommendation Workflow},
  author={Kalakoti, Yogesh and Yadav, Shashank and Sundar, Durai},
  journal={ACS Omega},
  year={2022},
  publisher={ACS Publications}
}

@article{johnson2010hidden,
  title={Hidden Markov model speed heuristic and iterative HMM search procedure},
  author={Johnson, L Steven and Eddy, Sean R and Portugaly, Elon},
  journal={BMC bioinformatics},
  volume={11},
  number={1},
  pages={1--8},
  year={2010},
  publisher={Springer}
}

@inproceedings{yuan2016lambdafm,
  title={Lambdafm: learning optimal ranking with factorization machines using lambda surrogates},
  author={Yuan, Fajie and Guo, Guibing and Jose, Joemon M and Chen, Long and Yu, Haitao and Zhang, Weinan},
  booktitle={Proceedings of the 25th ACM international on conference on information and knowledge management},
  pages={227--236},
  year={2016}
}

@article{blumer1987occam,
  title={Occam's razor},
  author={Blumer, Anselm and Ehrenfeucht, Andrzej and Haussler, David and Warmuth, Manfred K},
  journal={Information processing letters},
  volume={24},
  number={6},
  pages={377--380},
  year={1987},
  publisher={Elsevier}
}

@article{bank1971protein,
  title={Protein data bank},
  author={Bank, Protein Data},
  journal={Nature New Biol},
  volume={233},
  pages={223},
  year={1971}
}

@article{mcarthur2013comprehensive,
  title={The comprehensive antibiotic resistance database},
  author={McArthur, Andrew G and Waglechner, Nicholas and Nizam, Fazmin and Yan, Austin and Azad, Marisa A and Baylay, Alison J and Bhullar, Kirandeep and Canova, Marc J and De Pascale, Gianfranco and Ejim, Linda and others},
  journal={Antimicrobial agents and chemotherapy},
  volume={57},
  number={7},
  pages={3348--3357},
  year={2013},
  publisher={Am Soc Microbiol}
}

@article{li2006cd,
  title={Cd-hit: a fast program for clustering and comparing large sets of protein or nucleotide sequences},
  author={Li, Weizhong and Godzik, Adam},
  journal={Bioinformatics},
  volume={22},
  number={13},
  pages={1658--1659},
  year={2006},
  publisher={Oxford University Press}
}

@article{johnson2019billion,
  title={Billion-scale similarity search with {GPUs}},
  author={Johnson, Jeff and Douze, Matthijs and J{\'e}gou, Herv{\'e}},
  journal={IEEE Transactions on Big Data},
  volume={7},
  number={3},
  pages={535--547},
  year={2019},
  publisher={IEEE}
}

@article{FAMSA,
   author = {Deorowicz, Sebastian and Debudaj-Grabysz, Agnieszka and Gudyś, Adam},
   title = {FAMSA: Fast and accurate multiple sequence alignment of huge protein families},
   journal = {Scientific Reports},
   volume = {6},
   number = {1},
   pages = {33964},
   ISSN = {2045-2322},
   DOI = {10.1038/srep33964},
   url = {https://doi.org/10.1038/srep33964},
   year = {2016},
   type = {Journal Article}
}



\section*{Checklist}

\begin{enumerate}

\item For all authors...
\begin{enumerate}
  \item Do the main claims made in the abstract and introduction accurately reflect the paper's contributions and scope?
    \answerYes{}
  \item Did you describe the limitations of your work?
    \answerYes{}
  \item Did you discuss any potential negative societal impacts of your work?
    \answerNo{}
  \item Have you read the ethics review guidelines and ensured that your paper conforms to them?
    \answerYes{}
\end{enumerate}

\item If you are including theoretical results...
\begin{enumerate}
  \item Did you state the full set of assumptions of all theoretical results?
    \answerNA{}
        \item Did you include complete proofs of all theoretical results?
    \answerNA{}
\end{enumerate}

\item If you ran experiments...
\begin{enumerate}
  \item Did you include the code, data, and instructions needed to reproduce the main experimental results (either in the supplemental material or as a URL)?
    \answerYes{We will release code and datasets for reproducibility.}
  \item Did you specify all the training details (e.g., data splits, hyperparameters, how they were chosen)?
    \answerYes{See Sections~\ref{sec:contact} and~\ref{antibioticresistance}}
        \item Did you report error bars (e.g., with respect to the random seed after running experiments multiple times)?
    \answerYes{In figure~\ref{num_msa_graph}, we report the effect of different MSAs on the results.}
        \item Did you include the total amount of compute and the type of resources used (e.g., type of GPUs, internal cluster, or cloud provider)?
    \answerYes{See appendix~\ref{gpu_cluster}.}
\end{enumerate}

\item If you are using existing assets (e.g., code, data, models) or curating/releasing new assets...
\begin{enumerate}
  \item If your work uses existing assets, did you cite the creators?
    \answerYes{}
  \item Did you mention the license of the assets?
    \answerYes{}
  \item Did you include any new assets either in the supplemental material or as a URL?
    \answerYes{}
  \item Did you discuss whether and how consent was obtained from people whose data you're using/curating?
    \answerYes{}
  \item Did you discuss whether the data you are using/curating contains personally identifiable information or offensive content?
    \answerNA{}
\end{enumerate}

\item If you used crowdsourcing or conducted research with human subjects...
\begin{enumerate}
  \item Did you include the full text of instructions given to participants and screenshots, if applicable?
    \answerNA{}
  \item Did you describe any potential participant risks, with links to Institutional Review Board (IRB) approvals, if applicable?
    \answerNA{}
  \item Did you include the estimated hourly wage paid to participants and the total amount spent on participant compensation?
    \answerNA{}
\end{enumerate}

\end{enumerate}


\pagebreak
\appendix

\section{Appendix}

\subsection{Methodology Details.}
\label{methoddetails}

For protein-level prediction tasks, we extract representation features from the  [CLS] token of the last layer of ESM-1b and MSA-Transformer and add a new linear layer on top of it. Since there is no [CLS] token appended in AlphaFold, we average all residue  embeddings from the last  layer and add the same linear layer on top. For residue-level supervised prediction task (i.e. SS prediction), we add a linear layer on top of each residue embedding the last layer in all three PLMs. We train these models for each non-zero-shot task by fine-tuning all parameters, including both the new linear layer and the backbone representation model.

For contact prediction, we perform experiments differently for the three models given that the pairwise distance matrix of Evoformer can be directly extracted. For MSA-Transformer and ESM-1b, we follow the~\cite{rao2019evaluating}, extracting and normalizing the attention map from all hidden layers and then training a linear layer on these 2D maps and performing the regression task.


For  zero-shot fitness prediction,  MSAs are searched in the BFD dataset~\cite{jumper2021highly} using Jackhmmer~\cite{johnson2010hidden} with the default parameters. For annotation prediction tasks, we search MSAs from Uniref90~\cite{UniProt} database with Jackhmmer. For SS and contact prediction, we use the MSAs provided by ~\cite{rives2021biological}. 
MSAs are not used in the two supervised fitness prediction tasks because all sequences are highly similar with only a few  positions different. 
We report all results of Evoformer without template to avoid information leaking for the structure tasks.

\subsection{ESM-MSA details}
\label{ESM-MSAdetails}

\subsubsection{Training set construction}
ESM-MSA is essentially a two-tower-based network following sentence-BERT~\cite{reimers2019sentence} where each tower is represented by the ESM-1b encoder. Unlike sentence-BERT, we have developed an effective negative sampling method to choose more informative negative protein sequences rather than perform random sampling. The schematic of ESM-MSA is shown in Figure~\ref{esm1b_MSA} and more details are given below:

We first collect high-quality homologous sequence datasets. We use the public trRosetta training set\footnote{https://github.com/gjoni/trRosetta} as the ground truth data in this work.
Define $H=\{h_1,h_2,...,h_n\}$ as the entire protein sequence data set, where 
$h_{i}$ represents the aligned homologous sequences for protein $i$, usually called an MSA.
Define $h_{i}=\{q_1,q_2,...,q_n\}$ , where $q_{i}$ is an individual sequence in a same homologous family.
Then define  $D_{neg}=\{d_1,d_2,...,d_n\}$ as the data for negative sampling, where $d_i$ also denotes an individual protein sequence. 
We use Uniclust30\footnote{https://uniclust.mmseqs.com/} for sampling non-homologous sequences, which include 200 million individual protein sequences. 
We then construct a training set including homologous sequence pairs and non-homologous sequence pairs to train the model. Specifically, for a protein sequence $q_{i}$ belonging to  set $h_i$, we randomly sample its homologous sequence  $q_{pos} \in h_i$ from the same set 
\begin{figure}[H]
  \centering
  \includegraphics[width = 1\textwidth]{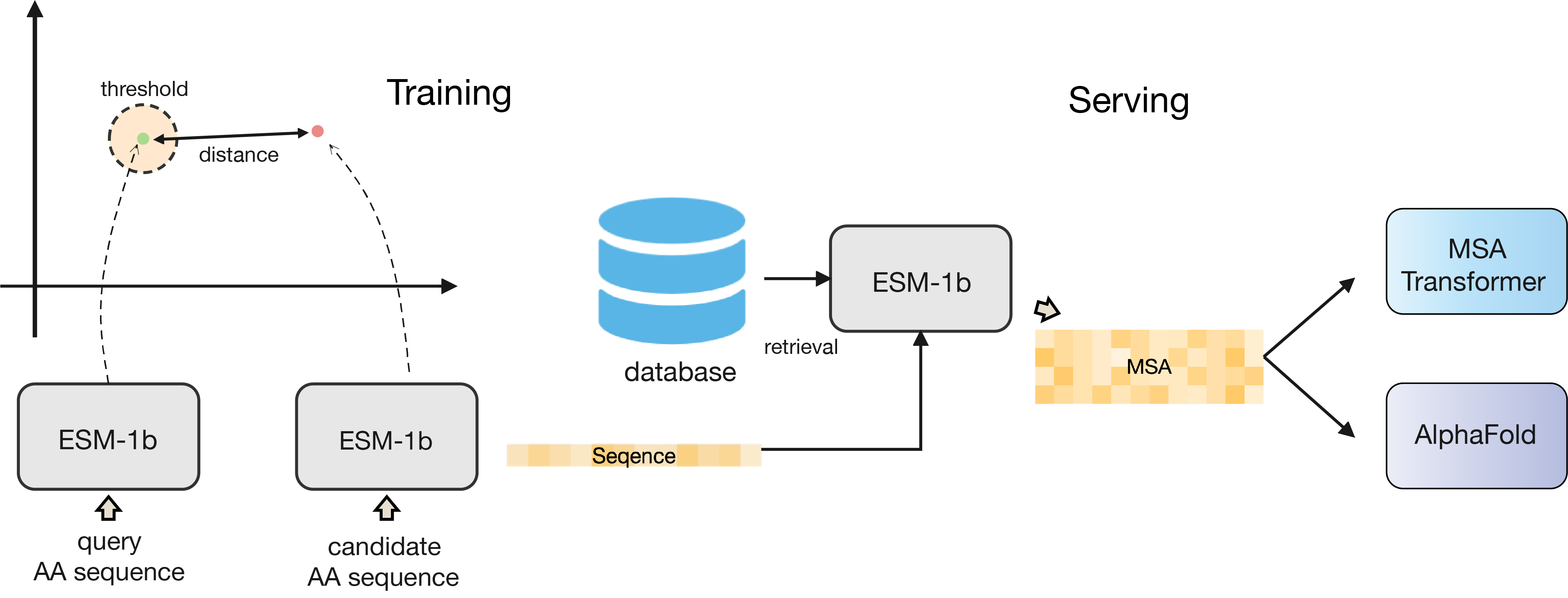}
  \caption{Training and serving schematic of ESM-MSA.}
  \label{esm1b_MSA}
\end{figure}
and calculate the biological identity $s \in [0,1]$ between the two sequences. Finally, we construct a positive sample pair in the training set $p^{pos}$ . Similarly, for the original sequence $q_{i} \in h_i$ , we sample a sequence  $q_{neg} \in D$ from the Uniclust30 database as a negative sample(non-homologous sequence), thus constructing a negative sample pair $P^{neg}$ .
We finally obtain the set of positive pairs $S_{pos}=\{ p_1^{pos},p_2^{pos},....,p_n^{pos}\}$ , and the set of negative pairs $S_{neg}=\{ p_1^{neg},p_2^{neg},....,p_m^{neg}\}$ , and then obtain the complete training set $S = S_{pos} \bigcup S_{neg}$ .

\subsubsection{Dynamic negative sampling}
In this paper, we use dynamic negative sampling at the training stage, a.k.a. hard negative sampling~\cite{yuan2016lambdafm}. That is, instead of randomly sampling sequences, we continuously select samples that are more difficult for the model to distinguish. By this way, the model convergence speed is greatly accelerated and the performance of the model is largely improved.
Specifically, for the newly initialized model, we randomly select a batch of samples from the database as negative samples for training. After that, in every new round, for each  sequence $q_i$, we randomly sample $n$ sequences from the database and calculate the Euclidean distance between the original sequence and sampled sequences by the model.
We select the sampled sequence with the closest Euclidean distance to the original sequence representation as the negative sample. Since the closest Euclidean distance means that the model is not able to distinguish it as negative.
To reduce the computational cost,  we use a sequence pooling approach, i.e., we first randomly sample $N$ sequences( $N \gg n$) from the database as a sub-database, and then perform subsequent sampling operations on such sub-database after the corresponding scores are computed by the model.

\subsubsection{Objective Function}
For the positive sample pair, the loss function is defined as follows:
$$Loss(p,q)_{pos} = (dist(p,q)-(1-s))^2$$
where $p$ is the original sequence, $q$ is the homologous sequence of $p$ , and $s$ is the biological identity of the two sequences. The definition is based on the hypothesis that the greater the biological sequence identity of the homologous sequences, the closer the Euclidean distance between the two should be through the representations given by the model. This means that between homologous sequences, the Euclidean distance also varies according to the biological sequence identity, and thus the model can obtain better generalization performance. Therefore, we used the mean square error loss function. Because $s \in [0,1]$, setting the distance threshold to 1 not only conforms to Occam's razor principle~\cite{blumer1987occam}, but also fits well with the loss function.

For the negative sample pair, the loss function is defined as follows:
$$Loss(p,q)_{neg} = -log z(p,q)$$
where $z(p,q)=\frac{e^{dist(p,1)}}{e^{dist(p,q)}+e^t}$  , $t$ is distance threshold. Actually, this is Cross Entropy function. For non-homologous sequences, we want the model to keep increasing the Euclidean distance between $p$ and $q$, so it is reasonable to use the distance threshold $t$ as a reference for the Cross Entropy, so that the Euclidean distance of the representations keeps moving away from the distance threshold.

\subsubsection{MSA Retrieval and Alignment}
First, we use the embedded vector of 'CLS' in  ESM-1b (trained by the above approach) as the protein representation. We calculate all protein sequences in
the database.
Then we use Faiss~\cite{johnson2019billion}, the library for quick embedding searching, to retrieve homologous sequences. Specifically, we calculate the Euclidean distance between sequence embeddings and keep those under the threshold as homologous to the query protein. One we obtain a set of homologous sequences, we use Famsa\cite{FAMSA} to efficiently align them and output an a3m format file.


\subsection{HHblits setting}
\label{hhblitsetting}
\subsubsection{Search the BFD database HHblits setting}
n\_iter: int = 3,
e\_value: float = 0.001,
maxseq: int = 1000000,
realign\_max: int = 100000,
maxfilt: int = 100000,
min\_prefilter\_hits: int = 1000,
all\_seqs: bool = False,
p: 20,
z: int = 500.

\subsubsection{Compare with the running time of ESM-MSA vs HHblits}
\label{speedmsa}
n\_iter: int = 1,
e\_value: float = 0.0001,
maxseq: int = 1000,
realign\_max: int = 100000,
maxfilt: int = 100000,
min\_prefilter\_hits: int = 1000,
all\_seqs: bool = False,
p: 20,
z: int = 500.

\begin{table}[H]
  \caption{HHblits speed compared with ESM-MSA (retieval \& alignment). The values means how many MSAs are searched in 12h with 12-core CPU.}
  \label{HHBLITS_with_esm_msa}
  \centering
  \begin{tabular}{lr}
    \toprule
    Model          & Number of MSAs  \\
    \midrule
    HHblits         & 1509  \\
    ESM-MSA           & \textbf{8800}  \\
    \bottomrule
  \end{tabular}
\end{table}
               
\subsection{Contact map Results on SCOPe}
\label{sec:contact}
See Table~\ref{cm_res} for details.
\begin{table}
  \caption{Contact map prediction, Precision@L, L/5, L/2.
}
  \label{cm_res}
  \centering
  \begin{tabular}{lrrr}
    \toprule
    Model         & Precision@$L$ &Precision@$L/2$ & Precision@$L/5$ \\
    \midrule
    ESM-1b    &0.540	&0.668&0.783   \\
    MSA Transformer & 0.660&0.784&0.872   \\
      Evoformer  &  \textbf{0.946}&\textbf{0.970}&\textbf{0.978}   \\
    \bottomrule
  \end{tabular}
\end{table}


\subsection{Antibiotic Resistance dataset detail}
\label{antibioticresistance}
The Antibiotic Resistance dataset is derived from experimentally verified bacterial antibiotic resistance proteins from the Comprehensive Antibiotic Resistance Database (CARD), and redundant sequences with 100\% identity are removed using the CD-HIT tool~\cite{li2006cd}. Finally, a total of 2602 protein sequences from 19 antibiotic classes are constructed for functional classification and analysis.

\subsection{GPU cluster}\label{gpu_cluster}
All our experiments are performed on the NVIDIA A40 with 48G GPU  memory.

\subsection{Remote Homology Detection (Evolutionary Understanding Task)}
 Here, we add the results of the remote homology detection task, which has exactly the same training and testing set in TAPE.  As shown in Table~\ref{Remote homology}, we can make the same observations as in the annotation prediction task. First, the performance of all three PLMs is substantially improved by pre-training, which shows the representation ability of them; Second, ESM-1b performs better than MSA-Transformer and Evoformer.  Our results are  consistent with those in TAPE. It is worth noting that although the homology detection task measures a model’s ability to detect structural similarity, it is essentially  formulated as a protein-level annotation or classification task, like the MIB and ABR tasks in this paper. By comparison, the typical  structural prediction tasks, including the SS prediction,  contact prediction, and 3D structure prediction, are atom- or residue-level prediction, where residue is often represented by the $C_{\alpha}$ or $C_{\beta}$ atom.  
\begin{table}[H]
  \caption{Results of the remote homology detection task.  `scratch' means random initialization for model parameters.}
  \label{Remote homology}
  \centering
  \begin{tabular}{lrr}
    \toprule
    Model          & Pretrained &Scratch  \\
    \midrule
    ESM-1b & \textbf{0.31} & 0.12\\
    MSA-Transformer & 0.22 & \textbf{0.13}\\
    Evoformer & 0.23 & 0.11 \\
    \bottomrule
  \end{tabular}
\end{table}


\end{document}